\documentclass[showpacs,preprintnumbers,prl,amsmath,amssymb,superscriptaddress, twocolumn, linenumbers]{revtex4}
\usepackage{graphicx}
\usepackage{endnotes}
\usepackage{bm}
\usepackage{epsfig}

\newcommand{\CeCoIn}{CeCoIn$_{5}$}

\begin{document}

%%%%%%%%%%%%%%%%%%%%% TITLE %%%%%%%%%%%%%%%%%%%%

\title{Anomalous Upper Critical Field in CeCoIn$_5$/YbCoIn$_5$ Superlattices
with a Rashba-type Heavy Fermion Interface}

%%%%%%%%%%%%%%%%%%%% AUTHORS %%%%%%%%%%%%%%%%%%

\author{S. K. Goh}
\affiliation{Department of Physics, Graduate School of Science, Kyoto University, Kyoto 606-8502, Japan}
\affiliation{Cavendish Laboratory, University of Cambridge, J. J. Thomson Avenue, Cambridge CB3 0HE, United Kingdom}

\author{Y. Mizukami}
\author{H. Shishido}
\altaffiliation{Current address: \textit{Department of Physics and Electronics, Osaka Prefecture University,
1-1 Gakuen-cho, Naka-ku, Sakai, Osaka 599-8531, Japan}}
\author{D. Watanabe}
\author{S. Yasumoto}
\author{M. Shimozawa}
\author{M. Yamashita}
\altaffiliation{Current address: \textit{RIKEN, Wako-shi, Saitama 351-0198, Japan}}
\affiliation{Department of Physics, Graduate School of Science, Kyoto University, Kyoto 606-8502, Japan}

\author{T. Terashima}
%\affiliation{Department of Physics, Graduate School of Science, Kyoto University, Kyoto 606-8502, Japan}
\affiliation{Research Center for Low Temperature and Materials Science, Kyoto University, Kyoto 606-8501, Japan}

\author{Y. Yanase}
\affiliation{Department of Physics, Niigata University, Niigata 950-2181, Japan}

\author{T. Shibauchi}
\affiliation{Department of Physics, Graduate School of Science, Kyoto University, Kyoto 606-8502, Japan}

\author{A. I. Buzdin}
\affiliation{Universit\'{e} Bordeaux I, LOMA, UMR 5798, F-33400 Talence, France}

\author{Y. Matsuda}
\affiliation{Department of Physics, Graduate School of Science, Kyoto University, Kyoto 606-8502, Japan}
\date{October 18, 2012}

%%%%%%%%%%%%%%%%%%% ABSTRACT %%%%%%%%%%%%%%%%%%%%

\begin{abstract}
We report a highly unusual angular variation of the upper critical field ($H_{c2}$) in epitaxial superlattices CeCoIn$_5$($n$)/YbCoIn$_5$(5), formed by alternating layers of $n$ and a $5$ unit-cell thick heavy-fermion superconductor CeCoIn$_5$ with a strong Pauli effect and normal metal YbCoIn$_5$, respectively. For the $n=3$ superlattice, $H_{c2}(\theta)$ changes smoothly as a function of the field angle $\theta$. However, close to the superconducting transition temperature, $H_{c2}(\theta)$ exhibits a cusp near the parallel field ($\theta=0^{\circ}$). This cusp behavior disappears for $n=4$ and $5$ superlattices. This sudden disappearance suggests the relative dominance of the orbital depairing effect in the $n=3$ superlattice, which may be due to the suppression of the Pauli effect in a system with local inversion symmetry breaking. Taking into account the temperature dependence of $H_{c2}(\theta)$ as well, our results suggest that some exotic superconducting states, including a helical superconducting state, might be realized at high magnetic fields. 

\end{abstract}

\pacs{74.25.Op, 81.15.Hi} 

%74.25.Op - superconductors - critical fields.
%74.70.Tx : Heavy-fermion superconductors
%81.15.Hi : Molecular, atomic, ion, and chemical beam epitaxy

\maketitle

%%%%%%%%%%%%%%%%%%%%% INTRO %%%%%%%%%%%%%%%%%%%%
%\section{Introduction}

In the absence of time reversal symmetry or space inversion symmetry, the Fermi surface (FS) can often be split into portions with different spin structures. To stabilize superconductivity under such conditions where spin degeneracy is lifted, unconventional pairing of quasiparticles is needed, leading to exotic superconducting states very different from the conventional BCS pairing state of ($\textbf{k}\uparrow$, $\textbf{-k}\downarrow$).
Considering the situation of the broken time reversal symmetry alone, Fulde and Ferrell \cite{Fulde64}, and Larkin and Ovchinnikov \cite{Larkin65} proposed the pairing state of ($\textbf{k}\uparrow$, $\textbf{-k+q}\downarrow$) on a Zeeman-split FS. This so-called FFLO pairing state leads to the modulation of the superconducting order parameter in real space with the modulation wavelength of the order of $1/|\textbf{q}|$. On the other hand, in the lack of space inversion symmetry, a Rashba-type spin-orbit coupling splits the FS into branches with spins of opposite rotation sense \cite{Rashba60}. When the magnetic field is applied to such a system, a pairing state with a finite center-of-mass momentum can also be realized, resulting in a helical superconducting state analogous to the FFLO phase.

However, such exotic superconducting states have been poorly explored because of the lack of suitable materials. Recent advancement in heavy fermion thin film fabrication technology \cite{Shishido10, Mizukami11} has enabled the preparation of superlattices formed by alternate stacking of c-axis oriented \CeCoIn\ and YbCoIn$_5$ with atomic layer thicknesses. The large Fermi velocity mismatch across the interface between \CeCoIn\ and YbCoIn$_5$ significantly reduces the transmission probability of quasiparticles, thereby ensuring quasi-two-dimensional superconductivity confined within \CeCoIn\ layers \cite{She12, Fenton85}. This provides a unique opportunity to explore the physics discussed above. This is because bulk \CeCoIn\ with strong Pauli effect has been reported to host the FFLO phase at low temperatures and high magnetic field \cite{Matsuda07, Radovan03, Bianchi03, Kumagai11}. In the superlattice, the electronic structure becomes two-dimensional, which is expected to stabilize the FFLO phase \cite{Matsuda07}. Moreover, the three-dimensional magnetic order \cite{Young07, Curro10, Kenzelmann08, Kenzelmann10}, which is responsible for the perplexing situation in the bulk sample, is expected to be strongly suppressed \cite{Shishido10} due to negligibly small RKKY interaction between the adjacent \CeCoIn\ block layers through the YbCoIn$_5$ spacer. Furthermore, the importance of \textit{local} inversion symmetry at the interface between \CeCoIn\ and YbCoIn$_5$ in the superlattice has recently been suggested to play a decisive role in superconducting properties, in particular when the thickness of \CeCoIn\ is only a few unit cells thick \cite{Maruyama12}.

In this Letter, we report the precise angular dependence of $H_{c2}$ of the \CeCoIn($n$)/YbCoIn$_5$(5) superlattices with $n=3, 4$ and $5$. We find that, for the $n=3$ superlattice, $H_{c2}(\theta)$ exhibits a highly unusual temperature evolution, which can not be explained for conventional thin films or superlattices. Using Ginzburg-Landau (GL) theory, we show that this is consistent with the scenario that the ultrathin, two-dimensional \CeCoIn\ enters an inhomogeneous phase at low temperatures, and the analysis of the $n$ dependence allows us to discover the importance of space inversion symmetry.
 
\begin{figure}[!t]\centering
      \resizebox{8.5cm}{!}{
              \includegraphics{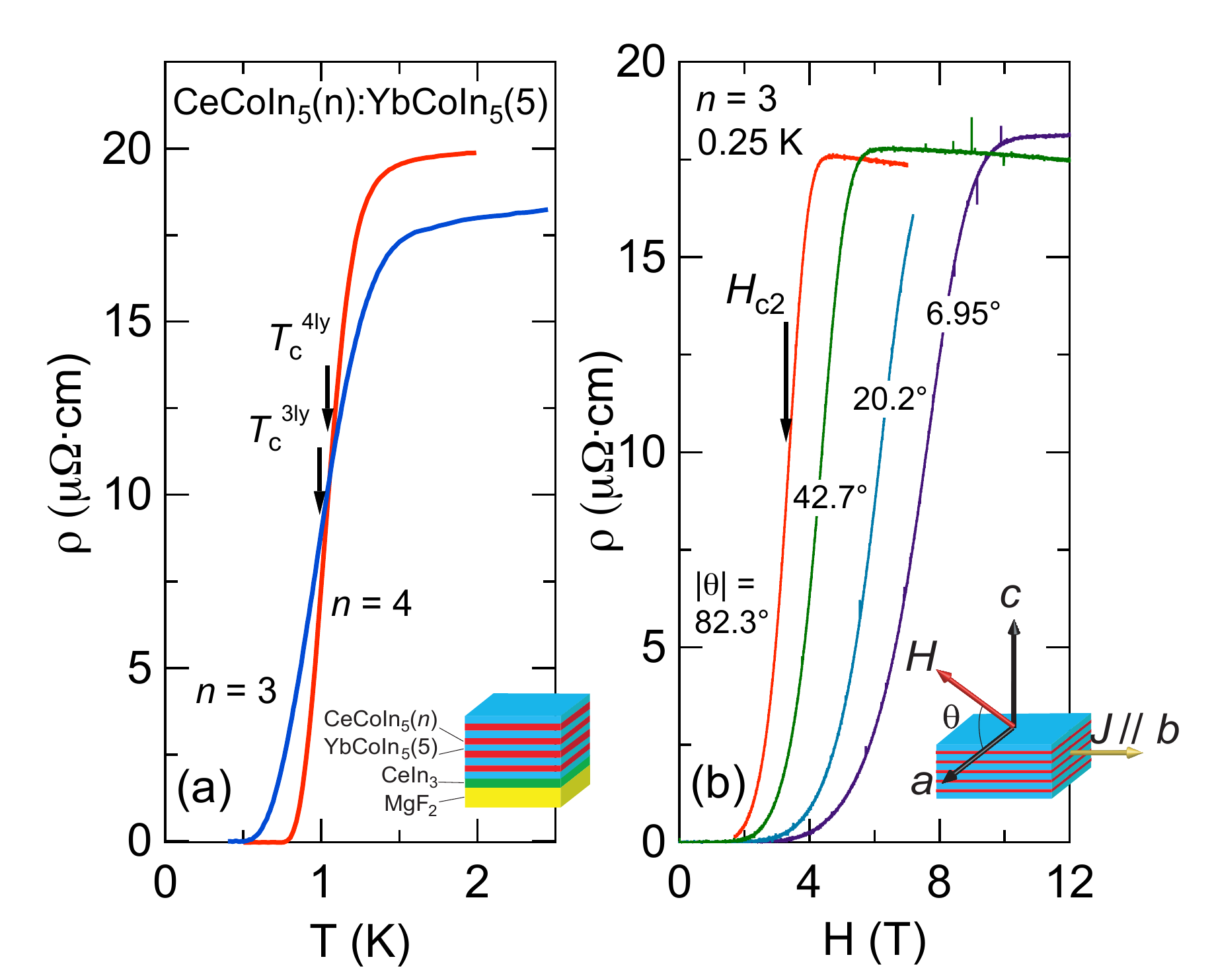}}                				
              \caption{\label{fig1} (Color online) (a) Temperature dependence of resistivity in $n=3$ and $n=4$ superlattices. $T_c$, defined as the temperature where the resistance is 50\% of the normal state resistance, is 0.99~K and 1.04~K for $n=3$ and $n=4$, respectively. The inset presents the schematic of the superlattice. (b) Field dependence of resistivity in the $n=3$ superlattice at 0.25~K at selected angles. $H_{c2}$ is similarly defined as the 50\% point of the normal state resistance. The inset shows the arrangement of the experimental configuration and the definition of $\theta$.}
\end{figure}

%%%%%%%%%%%%%%%%% EXPERIMENTAL %%%%%%%%%%%%%%%%%%%%
%\section{Experimental}
The \CeCoIn ($n$)/YbCoIn$_5$(5) superlattices used for this work were grown using the molecular beam epitaxial technique, where $n$ layers of \CeCoIn\ and $5$ layers of YbCoIn$_5$ were stacked alternately, typically repeated for 30 -- 60 times (Fig. \ref{fig1}a, inset). The growth details and the characterization of the superlattices are described elsewhere \cite{Mizukami11}. Resistivity measurements were performed using a standard four-contact method, with an electrical current of 3.16~$\mu$A flowing along the $b$-axis of \CeCoIn. The superlattice was mounted on the rotator of a dilution fridge, allowing the magnetic field direction to vary in the $ac$ plane, with $\theta=0^{\circ}$ corresponds to $H\parallel a$ (Fig. \ref{fig1}b, inset). In addition to the $n=3$ superlattice where $T_c=0.99$~K, we also measured $n=4$ and $n=5$ superlattices ($T_c=1.04$~K and $1.24$~K, respectively) for comparison.

%%%%%%%%%%%%%%%%%%%%% RESULTS %%%%%%%%%%%%%%%%%%
%\section{Results and Discussion}
%%%%%%%%%%%%%%%%Figure 1
%%%%%%%%%%%%%%%%%%%%

Fig. \ref{fig1}a shows the temperature dependence of resistivity for $n=3$ and $n=4$ superlattices, where zero resistivity is observed. Since $\xi_c(0)$, the zero temperature coherence length perpendicular to the conducting plane, is estimated to be about 3 unit-cells thick \cite{Matsuda07}, orbital depairing effect would ordinarily be weakened in these films. In this regard, it would also be interesting to study the $n=2$ superlattice. However, the $n=2$ system does not exhibit a full superconducting transition. When the magnetic field is applied, clear transition to the normal state is recorded, enabling an accurate determination of $H_{c2}$ as a function of angle (Fig. \ref{fig1}b).

%%%%%%%%%%%%%%%%Figure 2
\begin{figure}[!t]\centering
       \resizebox{8.5cm}{!}{
              \includegraphics{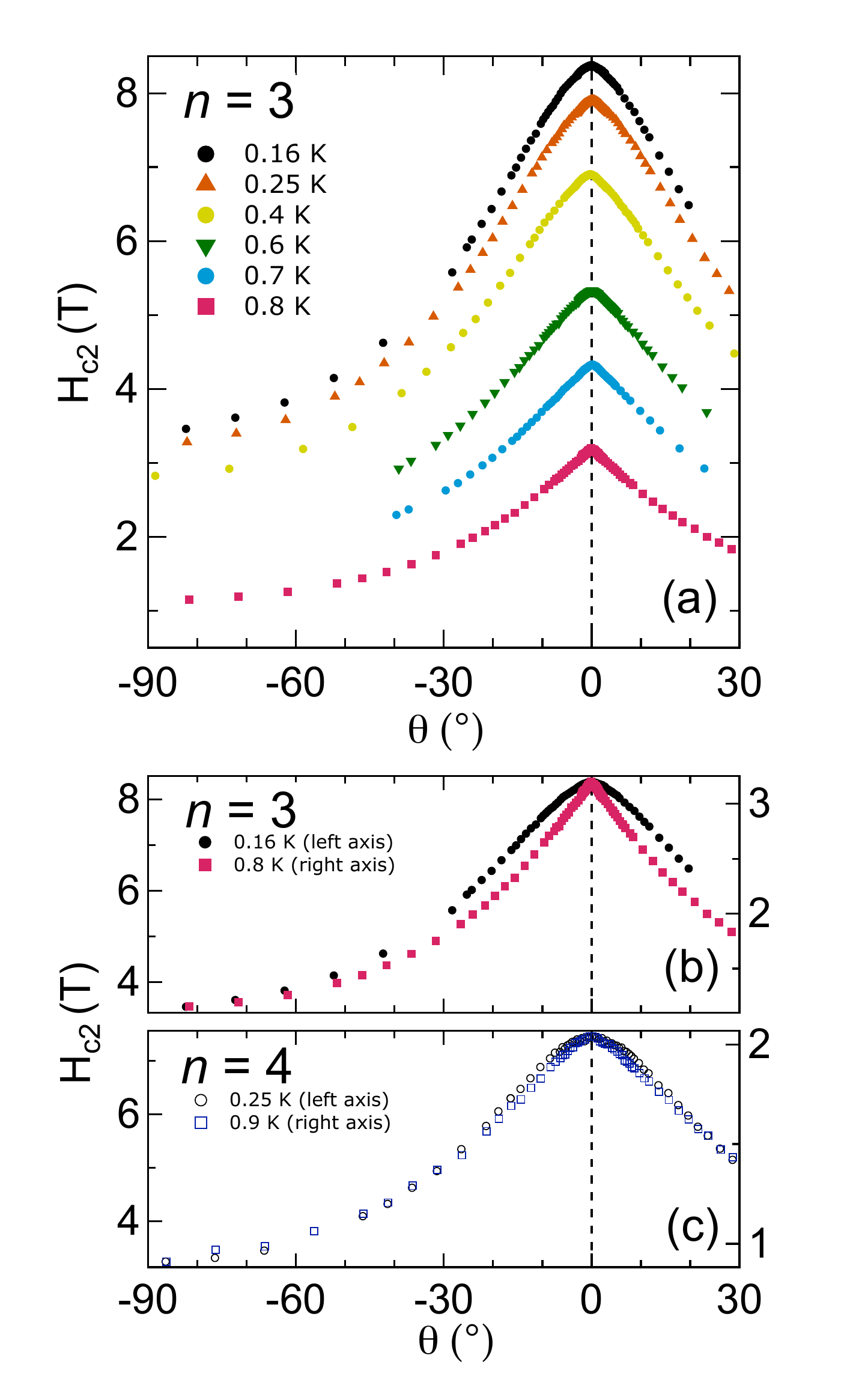}}                				
              \caption{\label{fig2} (Color online) (a) Angular dependence of $H_{c2}$ for the $n=3$ superlattice. (b) $H_{c2}(\theta)$ of the $n=3$ superlattice at 0.16~K and 0.8~K, normalised at $H_{c2}(0^{\circ})$ and $H_{c2}(90^{\circ})$, clearly exhibit different curvatures (c) For reference, $H_{c2}(\theta)$ of the $n=4$ superlattice at 0.25~K and 0.9~K, normalised at $H_{c2}(0^{\circ})$ and $H_{c2}(90^{\circ})$, are also plotted}
\end{figure}
%%%%%%%%%%%%%%%%%%%%

The angular dependence of $H_{c2}$ in an anisotropic bulk superconductor is smooth for all $\theta$, and can be described by \cite{Tinkhambook}
\begin{equation}
\left[\frac{H_{c2}(\theta)\rm{cos}\theta}{H_{c2}(0^{\circ})}\right]^2=-\left[\frac{H_{c2}(\theta) \rm{sin}\theta}{H_{c2}(90^{\circ})}\right]^2+1.
\label{eqn1}
\end{equation}
However, for a thin film with thickness $d$ smaller than $\xi_c$, $H_{c2}$ obeys the following equation first derived by Tinkham \cite{Tinkham63}:
\begin{equation}
\left[\frac{H_{c2}(\theta)\rm{cos}\theta}{H_{c2}(0^{\circ})}\right]^2=-\left|\frac{H_{c2}(\theta) \rm{sin}\theta}{H_{c2}(90^{\circ})}\right|+1.
\label{eqn2}
\end{equation}
Therefore, in the thin film limit $H_{c2}$ is non-differentiable at $0^{\circ}$ and follows a cusp-like dependence at small $\theta$.

We present the $H_{c2}(\theta)$ data of the $n=3$ superlattice in Fig. \ref{fig2}a. At 0.8~K, $H_{c2}(\theta)$ shows a kink at $0^{\circ}$, and the slope of $H_{c2}(\theta)$ increases monotonically as $|\theta|$ decreases. This is a characteristic behaviour predicted by Eq. (\ref{eqn2}). On the contrary, at 0.16~K, $H_{c2}(\theta)$ is a lot smoother at $0^{\circ}$, which is closer to the behaviour described by Eq. (\ref{eqn1}). For intermediate temperatures, the curvature of $H_{c2}(\theta)$ appears to fall in between the two limits governed by Eqs. (\ref{eqn1}) and (\ref{eqn2}), respectively. The distinct functional form of $H_{c2}(\theta)$ can be readily compared when we plot the datasets at 0.8~K and 0.16~K on the same axes, with normalised $H_{c2}$ at $0^{\circ}$ and $90^{\circ}$ (Fig. \ref{fig2}b). Note, however, that a similar exercise on the $H_{c2}(\theta)$ data of the $n=4$ superlattice does not reveal such a drastic difference (Fig. \ref{fig2}c). 
%%%%%%%%%%%%%%%%Figure 3
\begin{figure}[!t]\centering
       \resizebox{8.5cm}{!}{
              \includegraphics{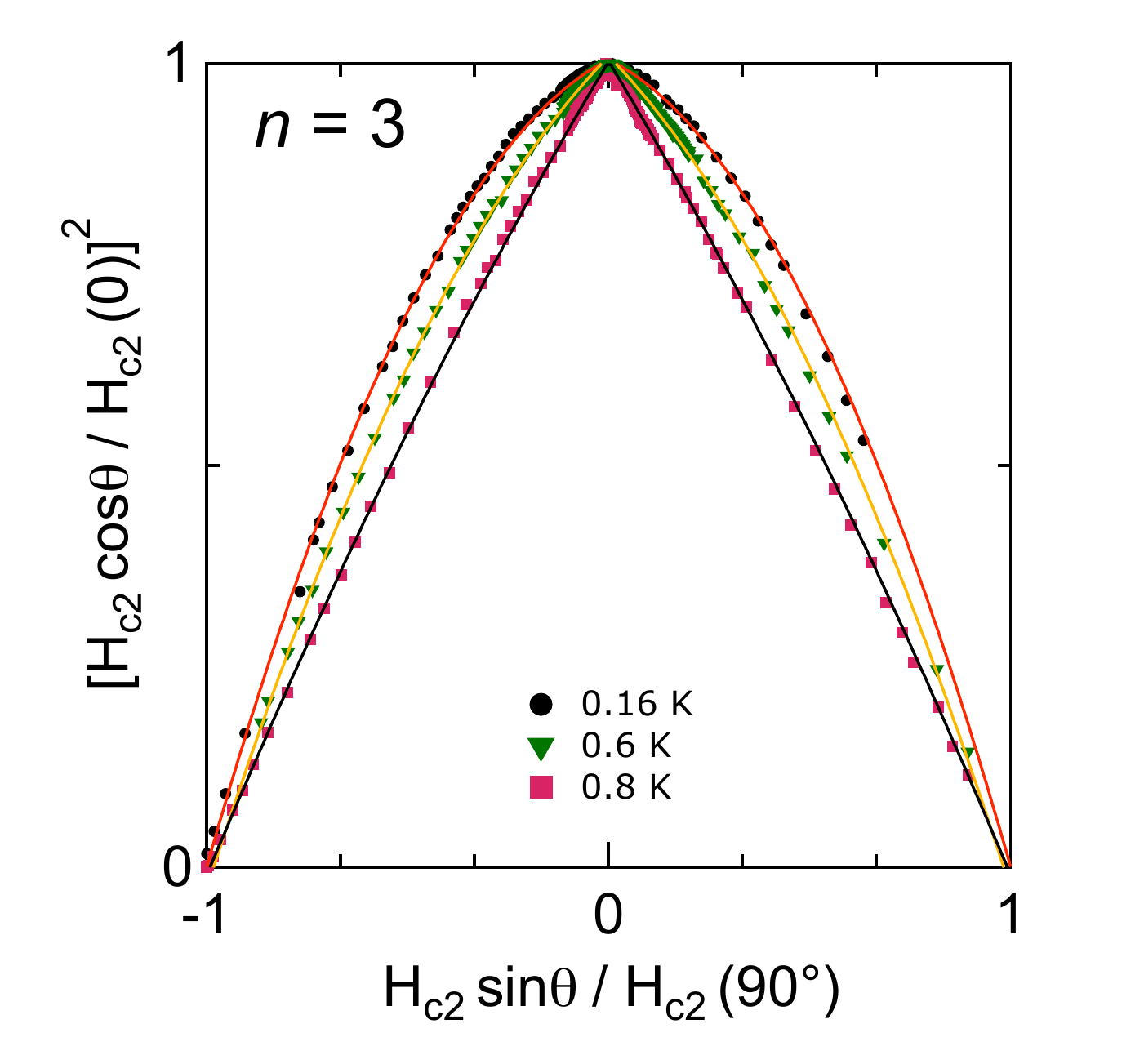}}                				
              \caption{\label{fig3} (Color online) The $H_{c2}(\theta)$ of the $n=3$ superlattice, replotted in an appropriate dimensionless form (see text). The solid lines are the fits to the data using the model described in the text.}
\end{figure}
%%%%%%%%%%%%%%%%%%%%

From the theoretical point of view, the existence of the FFLO phase can be discussed in the framework of GL theory by including the `gradient' term $k ( \left\vert \Pi _{x}\Psi \right\vert ^{2}+\left\vert \Pi
_{y}\Psi \right\vert ^{2})$ in the GL functional. In the FFLO state, $k$ becomes negative. Therefore the inclusion of the term lowers the total energy and guarantees the stability of the inhomogeneous phase. In the homogeneous BCS state, $k$ takes a positive value and it can be shown using the modified GL theory \cite{Buzdin97, Buzdin07} that for $k\geq0$ the slope of the critical field assumes the following form near $\theta=0^{\circ}$:

\begin{equation}
\left\vert\frac{\partial H_{c2}}{\partial \theta}\right\vert_{\theta=0^{\circ}}= kH_p\frac{8e}{a_0cg_a^2},
\label{eqnslope}
\end{equation}
where $H_p(T)$ is the line of the second order transition taking into account only the Pauli paramagnetic effect, $a_0$ is a numerical constant, and $g_a$ is related to the $g$ factor via $g=\sqrt{g_a^2\rm{cos}^2\theta+g_c^2\rm{sin}^2\theta}$. Therefore, the rounding of the cusp at $\theta=0^\circ$, which is manifested by the diminishing of $\left\vert\partial H_{c2}/\partial \theta\right\vert_{\theta=0^{\circ}}$, can come from two sources -- (i) a decreasing $k$ which accompanies the entrance to an inhomogeneous state, or (ii) the suppression of $H_p$ which signals the enhancement of the Pauli effect relative to the orbital effect.
%%%%%%%%%%%%%%%%Figure 4
\begin{figure}[!t]\centering
       \resizebox{8.5cm}{!}{
              \includegraphics{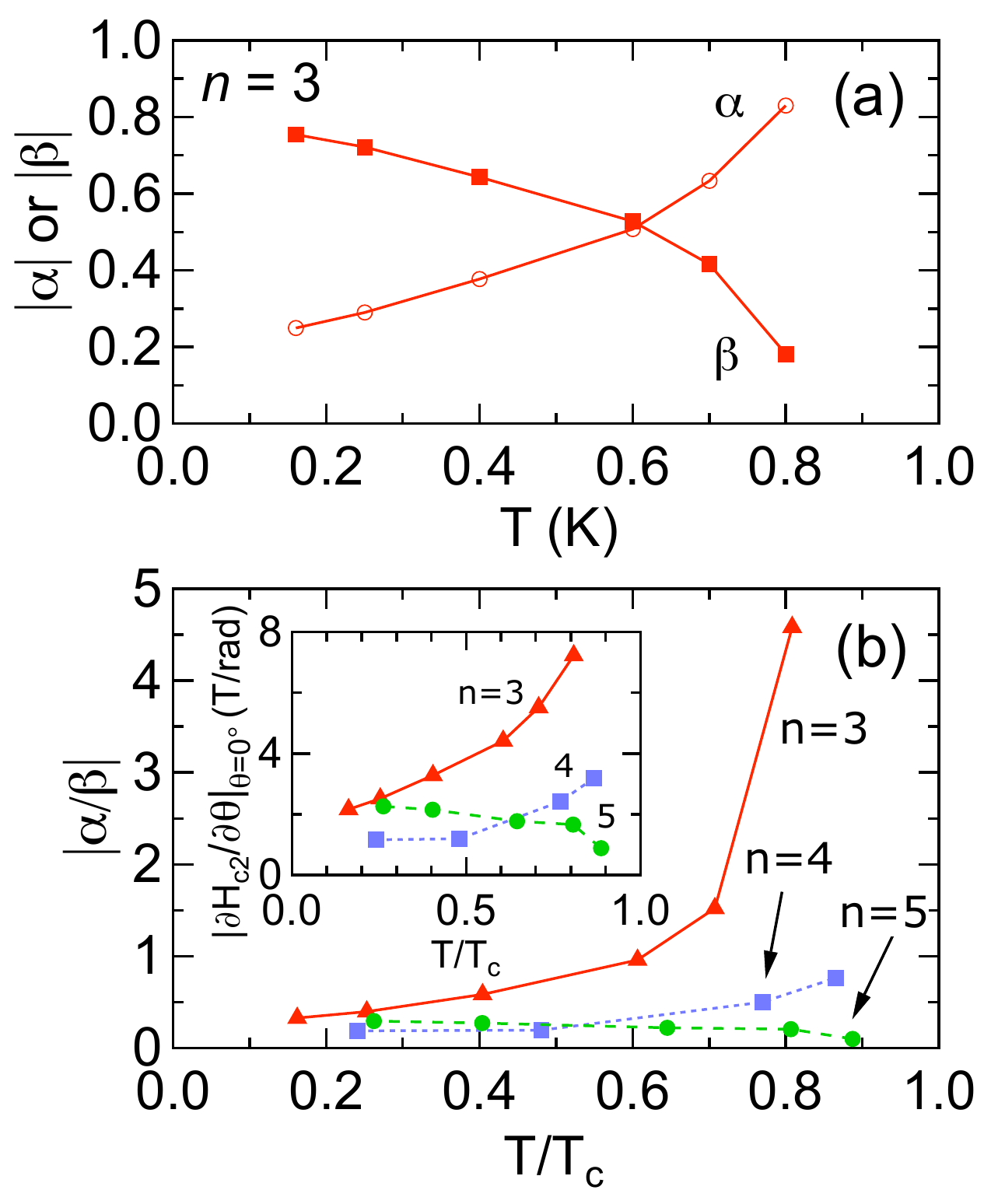}}                				
              \caption{\label{fig4} (Color online) (a) The contribution of the linear and quadratic components, characterized by $\alpha$ and $\beta$ respectively, as a function of temperature for the $n=3$ superlattice (b) The ratio of $\alpha/\beta$ for the $n=3$ superlattice as a function of the reduced temperature. The data for $n=4$ and $n=5$ superlattices are also shown for comparison. Inset: The temperature dependence of $\left\vert\partial H_{c2}/\partial \theta\right\vert_{\theta=0^{\circ}}$ extracted experimentally (see text). }
\end{figure}
%%%%%%%%%%%%%%%%%%%%

We now attempt to quantify the relative merits of Eqs. (\ref{eqn1}) and (\ref{eqn2}) in describing our $H_{c2}$ data. We replot the data in Fig. \ref{fig2} in the `dimensionless' form, as shown in Fig. \ref{fig3}. Using this representation, the graph would be linear for a pure Tinkham behaviour (Eq. \ref{eqn2}) and quadratic for the case of a simple, anisotropic bulk superconductor (Eq. \ref{eqn1}). However, it is apparent that the overall angular variation of $H_{c2}$ follows an intermediate behaviour. To move forward, we propose a description of our data using the model below:
\begin{equation}
\left[\frac{H_{c2}(\theta)\rm{cos}\theta}{H_{c2}(0^{\circ})}\right]^2=\alpha\left|\frac{H_{c2}(\theta) \rm{sin}\theta}{H_{c2}(90^{\circ})}\right|+\beta\left[\frac{H_{c2}(\theta) \rm{sin}\theta}{H_{c2}(90^{\circ})}\right]^2+1.
\label{eqn4}
\end{equation}
With this model, which essentially quantifies the relative contributions of Eqs. (\ref{eqn1}) and (\ref{eqn2}) through the coefficients $\alpha$ and $\beta$, an excellent description of the $H_{c2}(\theta)$ is achieved, as evidenced by the solid lines in Fig. {\ref{fig3}}.

The coefficients $\alpha$ and $\beta$ extracted using Eq. (\ref{eqn4}) are plotted in Fig. \ref{fig4}a for the $n=3$ superlattice. We find that $|\alpha|$ decreases as the temperature is lowered, whereas $|\beta|$ shows an opposite trend. We note that $|\alpha|$ and $|\beta|$ must sum to 1, as required by the construction of Eq. (\ref{eqn4}). This analysis confirms our earlier visual inspection that $H_{c2}(\theta)$ follows the Tinkham-like behaviour closer at high temperatures. To gauge the relative contribution of $\alpha$ and $\beta$, we calculate the ratio $|\alpha/\beta|$. As illustrated in Fig. \ref{fig4}b, for the $n=3$ superlattice, this ratio decreases drastically when the temperature decreases. Hence, at low temperatures, the squared term described by $\beta$ overwhelms the linear term associated with $\alpha$. The contrasting behaviour of the $n=4$ and $n=5$ superlattices is consistent with a much weaker temperature variation of $|\alpha/\beta|$.

%%%%%%%%%%%%%%%%Figure 5
\begin{figure}[!t]\centering
       \resizebox{8.5cm}{!}{
              \includegraphics{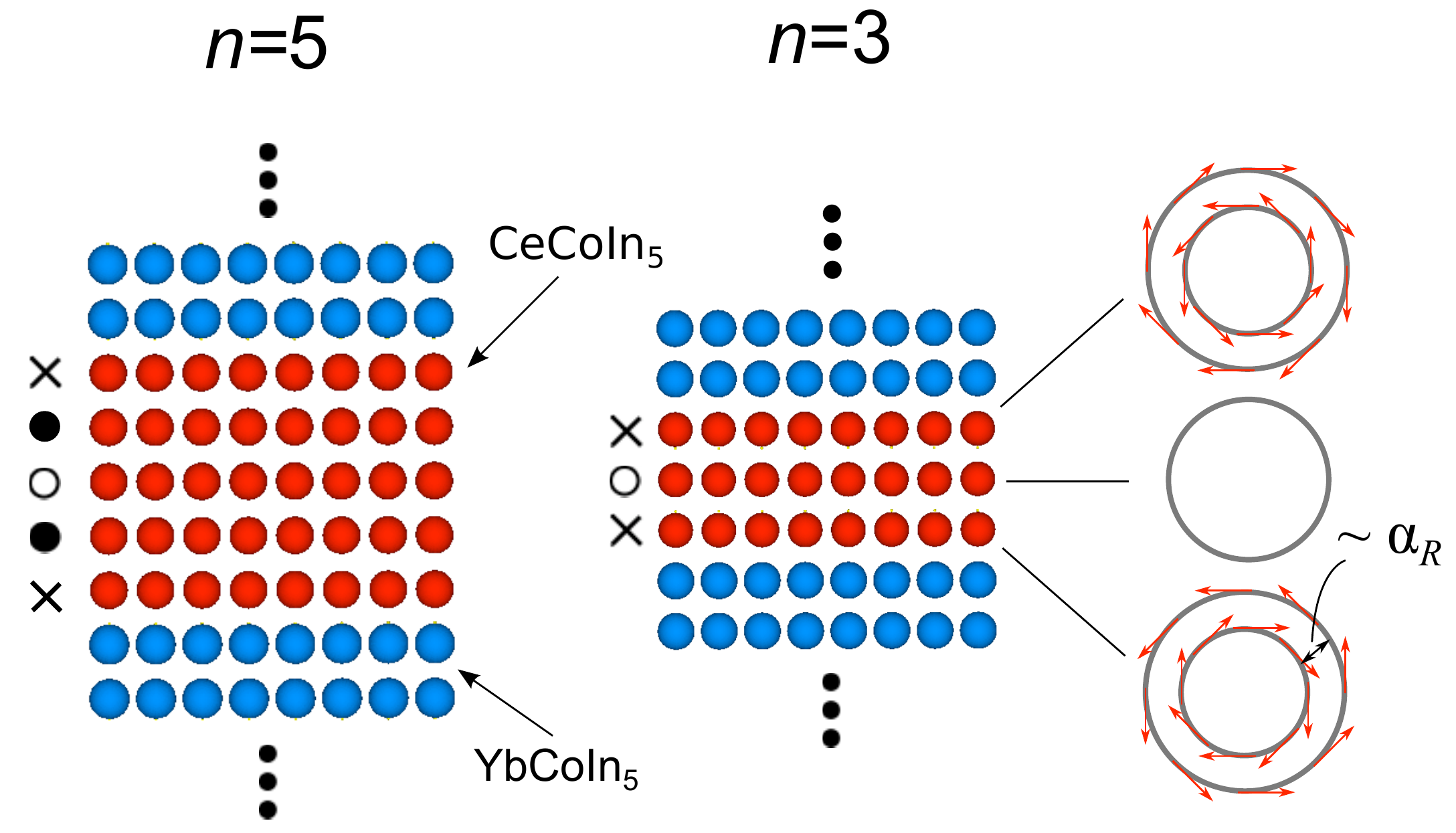}}                				
              \caption{\label{fig5} (Color online) Schematic diagrams showing the importance of local inversion symmetry breaking with reducing $n$. The \CeCoIn\ layers, shown in red, which possess (break) the local inversion symmetry are denoted by the open circles (crosses). With broken inversion symmetry, the FS is split into branches with spins of opposite rotation sense, and the degree of splitting is characterized by $\alpha_R$. Note that some inner layers, e.g. the layers indicated by the closed circles for $n=5$, also experience the effect of broken inversion symmetry, albeit with a much weaker effect and hence a much reduced $\alpha_R$. }
\end{figure}
%%%%%%%%%%%%%%%%%%%%

The magnitude of $\alpha$ is directly proportional to $\left\vert\partial H_{c2}/\partial \theta\right\vert_{\theta=0^{\circ}}$, since in the presence of the cusp both $|\alpha|$ and $\left\vert\partial H_{c2}/\partial \theta\right\vert_{\theta=0^{\circ}}$ are nonzero. Differentiating Eq. (\ref{eqn4}), we obtain $\left\vert\partial H_{c2}/\partial \theta\right\vert_{\theta=0^{\circ}}=(|\alpha|/2)[H_{c2}(0^\circ)^2/H_{c2}(90^\circ)]$, where all parameters can be extracted experimentally. 
The temperature dependence of $\left\vert\partial H_{c2}/\partial \theta\right\vert_{\theta=0^{\circ}}$ is plotted in the inset of Fig. \ref{fig4}b for all three superlattices.
For a thin film with a strong Pauli paramagnetic effect, the temperature dependence of $\left\vert\partial H_{c2}/\partial \theta\right\vert_{\theta=0^{\circ}}$ for the $n=5$ superlattice can be expected, i.e. a small but increasing slope as the temperature is lowered, probably dominated by an increasing $H_p$ with decreasing temperature. However, for $n=4$ and $n=3$ superlattices, in particular $n=3$, the behaviour is strikingly different in that $\left\vert\partial H_{c2}/\partial \theta\right\vert_{\theta=0^{\circ}}$ is a lot larger near $T_c$ and shows an opposite trend in the temperature evolution. Since $H_p(T)$ should increase upon cooling, the suppression of $\left\vert\partial H_{c2}/\partial \theta\right\vert_{\theta=0^{\circ}}$ at low temperatures can only be attributed to a rapid decrease of $k$. Our results therefore provide evidence that \textit{an FFLO-like inhomogeneous state is approached} at low temperatures.

We now provide a mechanism to explain the enhancement of $\left\vert\partial H_{c2}/\partial \theta\right\vert_{\theta=0^{\circ}}$ as the thickness of the film is reduced. In view of the fact that $\xi_c$ diverges near $T_c$, this exceptional sensitivity of $\left\vert\partial H_{c2}/\partial \theta\right\vert_{\theta=0^{\circ}}$ to layer thickness $d$ is surprising, since $d\ll\xi_c$ is satisfied for all three superlattices studied near $T_c$. Instead, it appears that the orbital pair breaking is a lot more effective in the $n=3$ superlattice  \cite{Note1}. To explain our data, we point out the importance of space inversion symmetry. In a system where space inversion symmetry is broken, it has been shown that the Pauli effect can be substantially reduced for all magnetic field orientations \cite{Mineev11}. Here, the inversion symmetry is locally broken at the top and the bottom `interface' \CeCoIn\ layers at the immediate proximity to YbCoIn$_5$ block layers. Although the inner layers might also break the local inversion symmetry, the effect has been theoretically shown to be much weaker \cite{Maruyama12}. Thus, with the reduction of $n$, the fraction of the non-centrosymmetric interface layers increases rapidly (Fig. \ref{fig5}). We argue that local inversion symmetry breaking is responsible for the enhancement of $\left\vert\partial H_{c2}/\partial \theta\right\vert_{\theta=0^{\circ}}$ with decreasing $n$ through the weakening of the Pauli effect. 

In the presence of the local inversion symmetry breaking, coupled with the fact that cerium has a large atomic number, the Rashba-type spin-orbit coupling is expected to be strong \cite{Rashba60}, giving rise to FS branches where the degree of splitting is proportional to the coupling strength $\alpha_R$ (Fig. \ref{fig5}). In this case, because of the violation of spin and momentum conservations along with the large effective mass, the coupling between a non-centrosymmetric layer with a Rashba-split FS and a centrosymmetic layer where the FS is intact is expected to be weak, leading to an overall reduction of the interlayer hopping integral $t_c$. In the limit where $\alpha_R\gg t_c$, interesting physics such as the enhancement of spin susceptibility have been predicted by Ref. \cite{Maruyama12}. Therefore, by combining the prospect of manipulating both the time reversal and space inversion symmetries, these superlattices offer a new avenue for detailed study of exotic superconducting phases, e.g. a helical superconducting phase \cite{Kaur05}, and novel vortex physics \cite{Hiasa09}.

In summary, we have measured and analysed $H_{c2}(\theta)$ of \CeCoIn($n$)/YbCoIn$_5$(5) superlattices for $n=3, 4$ and 5. We have shown that the rounding of the cusp at $\theta=0^\circ$ at low temperatures can be interpreted as a signature that the system approaches an FFLO-like phase. Furthermore, the drastic difference in the temperature dependence of $H_{c2}(\theta)$ for different $n$ can be understood by considering the importance of local inversion symmetry breaking, the effect of which is much stronger in the $n=3$ superlattice. These artificial heavy fermion superlattices where the effects of Pauli paramagnetism and Rashba interaction entangle thus offer a new playground for exploring exotic superconducting phases.

\textbf{Acknowledgement.} The authors acknowledge fruitful discussions with V. Mineev, R. Ikeda, S. Fujimoto and A. Balatsky. This work was supported by Grant-in-Aid for the Global COE program ``The
Next Generation of Physics, Spun from Universality and Emergence",
Grant-in-Aid for Scientific Research on Innovative Areas ``Heavy
Electrons'' (No. 20102006, 23102713) from MEXT of Japan, and
KAKENHI from the JSPS. S. K. G. acknowledges the JSPS for a fellowship. A. I. B. thanks Kyoto University for hospitality.

%%%%%%%%%%%%%%%%%% BIBLIOGRAPHY USING BIBTEX%%%%%%%%%%%%%%%%%%%%
%\bibliographystyle{phjcp.bst}
%\bibliography{AllRefs.bib}

%%%%%%%%%%%%%%%%%% END OF BIBLIOGRAPHY USING BIBTEX%%%%%%%%%%%%%%%%%%%%

\end{document}